\documentclass{article}
\usepackage{spconf,amsmath,graphicx}

\usepackage{booktabs} % for professional tables
\usepackage{bm}
\usepackage{amsmath}
\usepackage{amssymb}
\usepackage{url}
\usepackage{multirow}

\title{Self-Supervised Learning for speech recognition with Intermediate layer supervision }
\name{\begin{tabular}{c} Chengyi Wang$^{1,2}$ \thanks{Work done during an internship at Microsoft.},  Yu Wu$^2$, Sanyuan Chen$^2$, Shujie Liu$^2$, Jinyu Li$^2$,  Yao Qian$^2$,   Zhenglu Yang$^1$ \end{tabular} }

\address{$^1$ NanKai University, China, $^2$Microsoft Corporation}
\begin{document}
%\ninept
%
\maketitle
\begin{abstract}
 Recently, pioneer work finds that speech pre-trained  models can solve full stack speech processing tasks, because the model utilizes bottom layers to learn speaker-related information and top layers to encode content-related information.   Since the network capacity is limited, we believe the speech recognition performance could be further improved if the model is dedicated to audio content information learning. To this end, we propose {\bf{I}}ntermediate {\bf{L}}ayer {\bf{S}}upervision for {\bf{S}}elf-{\bf{S}}upervised {\bf{L}}earning (ILS-SSL), which forces the model to concentrate on content information as much as possible by adding an additional SSL loss on the intermediate layers.  Experiments on LibriSpeech test-other set show that our method outperforms HuBERT significantly, which achieves a 23.5$\%$/11.6$\%$ relative word error rate reduction in the w/o language model setting for base/large models.  Detailed analysis shows the bottom layers of our model have a better correlation with phonetic units, which is consistent with our intuition and explains the success of our method for ASR. We will release our code and model at \url{https://github.com/microsoft/UniSpeech}.
\end{abstract}
\begin{keywords}
Self-supervised learning, Automatic speech recognition
\end{keywords}
\section{Introduction}
\label{sec:intro}
Automatic speech recognition (ASR) has seen rapid improvement over the last few years due to the advances in deep learning. However, neural networks benefit from large quantities of labeled training data, which is much harder to come by than unlabeled data. This drives research in self-supervised learning (SSL), which attempts to learn representations from audio alone with some pretext tasks and then fine-tune the model on the supervised data.  Examples of such pretext tasks include generative task 
\cite{DBLP:conf/interspeech/ChungHTG19,DBLP:conf/icassp/LiuYCHL20,DBLP:journals/taslp/LiuLL21,DBLP:journals/corr/abs-2012-06659,chung2020generative,wang2020unsupervised}, discriminative task \cite{oord2018representation,DBLP:conf/interspeech/SchneiderBCA19,kharitonov2021data,DBLP:conf/iclr/BaevskiSA20,DBLP:conf/nips/BaevskiZMA20,zhang2020pushing,hubert} or multi-task \cite{DBLP:conf/interspeech/PascualRSBB19,DBLP:conf/icassp/RavanelliZPSMTB20,w2v_bert,DBLP:conf/icml/0002WQK0WZ021}. % prediction of future or masked audio features \cite{} or distinguish near-by or masked features from negative samples \cite{DBLP:conf/nips/BaevskiZMA20}. %Recent work show that self-supervised pre-training allow models perform well on ASR tasks. For instance, wav2vec 2.0 \cite{DBLP:conf/nips/BaevskiZMA20} and HuBERT \cite{} demonstrate remarkable performances with 60k hours of unpaired data and 1 hour labeled data on Librispeech.

To perform well on the downstream ASR task, the SSL objectives need to be well designed to learn spoken content information. Some successful SSL methods show that the learnt representations are highly correlated with phonetic units, such as wav2vec 2.0 \cite{DBLP:conf/nips/BaevskiZMA20} and HuBERT \cite{hubert}. However, studies indicate that not all layers have such a high correlation and different layers can learn disentangled aspects of speech.  \cite{DBLP:journals/corr/abs-2105-11084} train a supervised phoneme classifier on the top of frozen representations of each of the 24 blocks from a wav2vec 2.0 \textsc{Large} model and find that blocks 10-21 provide much lower phoneme error rate than others. In HuBERT \cite{hubert}, the authors run k-means clustering on representations of each of 12 blocks of a \textsc{Base} model and show that cluster assignments from blocks 5-12 have higher mutual information score with force-aligned phonetic transcripts. Furthermore, \cite{superb} collect representations from different HuBERT layers and weighted sum them for various downstream tasks where the weights can be learned. The model tends to assign larger weights to top layers for phoneme recognition task, while assign larger weights to bottom layers for speaker-related tasks. These observations indicate that the middle to top layers are better at learning content knowledge than the lower layers. 

Since a network capacity is limited, an interesting question is  whether we can free up its capacity from learning speaker characteristic and instead forcing more layers to learn content information. Following this motivation, we propose {\bf{I}}ntermediate {\bf{L}}ayer {\bf{S}}upervision for {\bf{S}}elf-{\bf{S}}upervised {\bf{L}}earning (ILS-SSL). Specifically, instead of computing the self-supervised loss only on the top layer, we also compute the loss on the intermediate layers. In this way, the lower layers must learn more content information to optimize the intermediate SSL loss. In this work, we use the masked prediction loss proposed in HuBERT \cite{hubert} as our objective since this loss can help the model learn phonetic information and perform well on ASR task. 

Our experiments either use the LibriSpeech 960h dataset or the Libri-Light 60k dataset for pre-training. In the \textsc{Base} model setting (pre-training with LibriSpeech 960h), our models outperform baselines on all fine-tuning subsets (1h, 10h and 100h). Among them, the model fine-tuned with the 100h subset outperforms the competitive HuBERT baseline by a 25.4$\%$/23.5$\%$ relative word error rate (WER) reduction on the test-clean/test-other testsets when decoding without a language model. When combining with an external 4-gram language model, our model still outperforms HuBERT by relative  11.7$\%$/14.8$\%$ WER.  In the \textsc{Large} model setting (pre-training with Libri-Light 60kh),  our method shows relative 9.5$\%$/11.6$\%$ WER reduction in the w/o language model setting when fine-tuning with 960h LibriSpeech data. When combining with Transformer language model, we reach a WER score of 1.8/3.2. We further conduct analysis to explain why the simple method could outperform baselines.  We find that ILS-SSL significantly improves the phonetic information learning for the bottom layers of the model. We also evaluate our model on the SUPERB benchmark \cite{superb} which includes ten different downstream tasks in four aspects of speech: content, speaker, semantics, and paralinguistics.  The results also indicate our model is good at content and semantic related tasks, and not good at speaker related tasks. The phenomenons underscore the fact that intermediate layer supervision could force the model to concentrate on content related information learning, leading to a better speech recognition performance.

\section{Background}
In this work, we apply our pre-training method on the HuBERT model, which benefits from an offline clustering step to provide target labels for a BERT-like prediction loss \cite{devlin2018bert}. The backbone includes a convolutional feature encoder and a Transformer \cite{vaswani2017attention} context encoder. During pre-training, the Transformer consumes corrupted features $\mathbf{\tilde{X}}$ and output hidden states $\mathbf{H}$ at the top layer. The network is optimized to predict the discrete target sequence $\bm{z}$, where each $z_t \in [C]$ is a $C$-class categorical variable.  The distribution over codewords is parameterized with 

\begin{equation}
    p(c|\mathbf{h}_t) = \frac{{\exp}({sim}(\mathbf{W}\mathbf{h}_t, \mathbf{e}_c)/\tau)}{\sum_{c'=1}^{C}{\exp}({sim}(\mathbf{W}\mathbf{h}_t, \mathbf{e}_{c'})/\tau)}
\label{eq1}
\end{equation} 
where $\mathbf{W}$ is a projection matrix, $\mathbf{h}_t$ is the hidden state for step $t$, $\mathbf{e}_c$ is the embedding for codeword $c$, sim($\mathbf{a}, \mathbf{b}$) computes the cosine similarity and $\tau=0.1$ scales the logit. A key ingredient of HuBERT is that the prediction loss is only applied over the masked regions, forcing the model to learn a combined acoustic and language model over the continuous inputs.

HuBERT adopts an iterative re-clustering and re-training process: For the first iteration, the discrete targets $\bm{z}$ are assigned by clustering on the MFCC features of the training data; For the second iteration, a new generation of training targets are created by clustering the learned latent representations from the first iteration model. The authors study how each layer performs when used to cluster and generate $\bm{z}$. They measure the correlation between generated discrete units and frame-level force-aligned phonetic units. For a 12-layers model, the 5th-8th layers of the first iteration model and the 5th-12th layers of the second iteration model have much higher mutual information score than others, indicating that the middle to top layers are better at phonetic information learning.

A concurrent work is w2v-BERT \cite{w2v_bert}, which bears similarity to ours. It directly optimizes a contrastive loss at an intermediate layer and a masked prediction loss at the top layer. The motivation of w2v-BERT is to use the codebook as pseudo-labels to simiplify training pipeline, while our work aims to force the model concentrate on ASR tasks. In addition, we  propose how to stabilize large-scale speech pre-training on GPUs, and  discuss the side-effect for some non-ASR tasks. 

\section{Method}
\label{sec:method}

\begin{figure}
\centering
  \label{model}
  \includegraphics[width=0.42\textwidth]{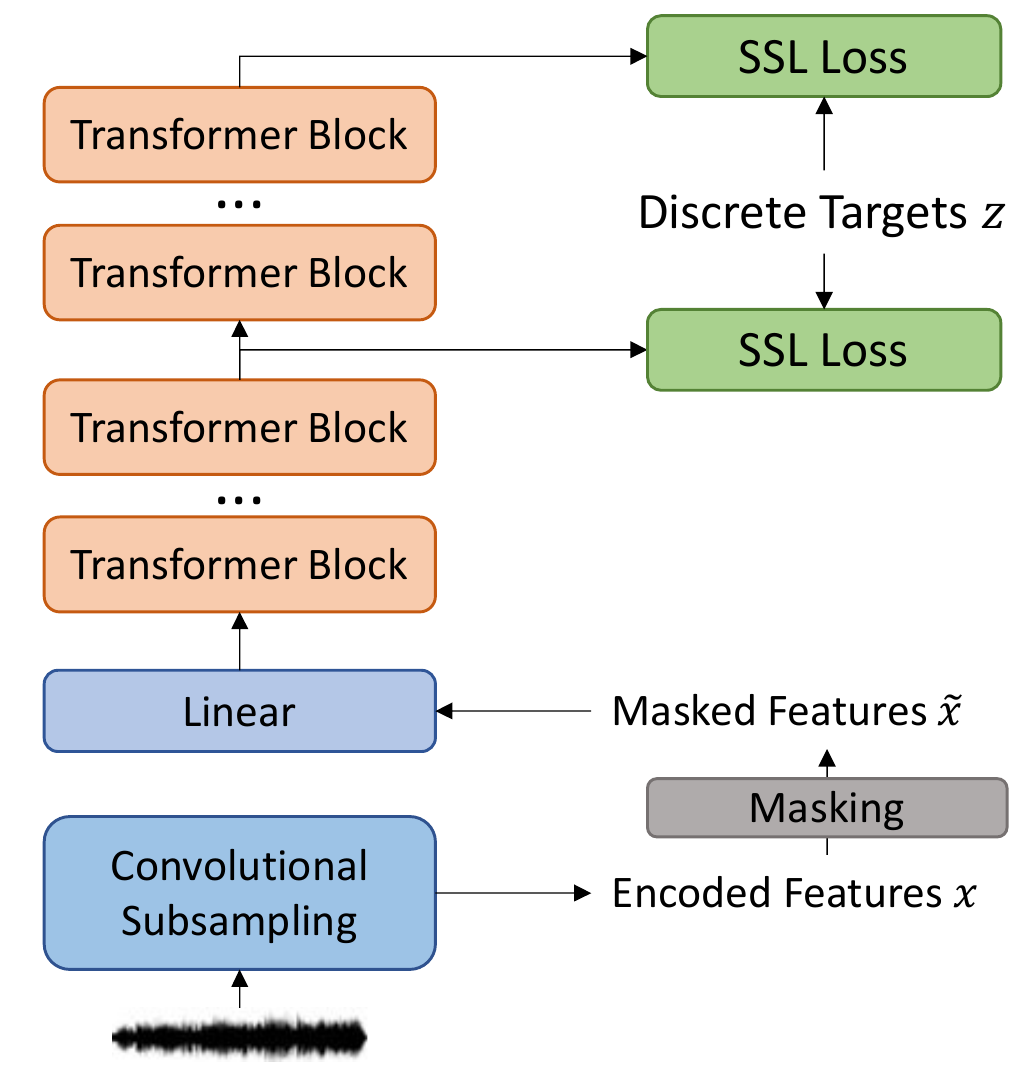}
  \caption{Model Architecture.}
\label{model}
\vskip -0.2in
\end{figure}
\subsection{Model Architecture}
%Our method is agnostic to the model architecture. To setup a fair comparison, we adopt the Transformer backbone used in wav2vec2.0 \cite{} and HuBERT \cite{}. 
Our model architecture backbone is the same as HuBERT. As shown in Figure \ref{model}, it contains a convolutional feature encoder and a Transformer encoder. The convolutional encoder is composed of seven blocks of temporal convolution followed by layer normalization and GELU activation layer. The temporal convolutions have 512 channels with strides (5,2,2,2,2,2,2) and kernel widths (10,3,3,3,3,2,2), resulting in each output representing about 25ms of audio strided by 20ms. Then we corrupt the output representations $\mathbf{X}$ by replacing part of features with a mask embedding and use this masked $\tilde{\mathbf{{X}}}$ as Transformer encoder input. We train Transformer in two different configurations: \textsc{Base} and \textsc{Large}. The \textsc{Base} model contains 12 layers with model dimension 768, inner dimension 3072 and 12 attention heads. The \textsc{Large} model contains 24 layers with dimension 1024, inner dimension 4096 and 16 attention heads. The Transformer encoder is equipped with a convolution based relative position embedding layer with kernel size 128 and 16 groups at bottom. 

Besides the convolutional position embedding, we also add a bucket relative position embedding which is encoded according to the offset between the ``key" and ``query" in the Transformer self-attention mechanism. We follow the implementation in \cite{DBLP:journals/jmlr/RaffelSRLNMZLL20}. Let $\{\mathbf{h}_i^{l-1}\}_{i=1}^{T}$ denote the input hidden states for the $l$-th self-attention layer, each $\mathbf{h}_i^{l-1}$ is linearly projected to a triple of query, key and value $(\mathbf{q}_i, \mathbf{k}_i, \mathbf{v_i})$ as:
\begin{equation}
    \mathbf{q}_i,\mathbf{k}_i,\mathbf{v}_i = \mathbf{h}_i^{l-1} \mathbf{W}^Q, \mathbf{h}_i^{l-1} \mathbf{W}^K, \mathbf{h}_i^{l-1} \mathbf{W}^V
\end{equation}
The self-attention outputs $\{\tilde{\mathbf{h}}_i^l\}_{i=1}^{T}$ are computed via:
\begin{align}
a_{ij} & \propto \exp (\frac{ \mathbf{q}_i \cdot \mathbf{k}_j }{\sqrt{d}} + r_{i-j}) \\
\tilde{\mathbf{h}}_i^{l} &= \sum_{j=1}^{T}{ a_{ij} \mathbf{v}_j }
\end{align}
where $r_{i-j}$ is a learnable scalar bias according to relative position distance $(i-j)$, $d$ is the encoder dimension. The embedding parameters are shared across all layers. In this work, we use 320 embeddings and each corresponds to a range of possible key-query offsets. The range increases logarithmically up to an offset of 800, beyond which we assign all relative offsets to the same embedding. We will analyze the effect of bucket relative position embedding in our experiments.
\vskip -0.2in

\subsection{Intermediate Layer Supervision}
The common practice of SSL is to compute the self-supervised loss on the top layer, such as wav2vec 2.0 and HuBERT. However, the lower layers of such a pre-trained model is shown to have a low correlation with phonetic information. In this work, we propose to apply intermediate layer supervision to encourage lower layers to learn content knowledge. During pre-training, we select a set of layers $K$ as supervised layers and compute the masked prediction loss on the output hidden states ${\mathbf{h}^l}$, where $l \in K$. The prediction targets are the same for every supervised layer. The final loss is defined as:
%also apply the loss on the middle layers. Specifically, once the feature encoder has transformed the input into latent representations $\bm{x}$, we randomly mask some regions and feed the masked sequence $\tilde{\bm{x}}$ to the Transformer encoder. Then we select a set of layers $K$ and use the layer representation to predict the k-means assignments:
\begin{equation}
    L  = -\sum_{l \in K}\sum_{t \in M}{\rm log}p^l(z_t|\mathbf{h}^l_t) % = \sum_{l \in K}\sum_{t \in M}{\rm log}p(z_t|{\tilde{x}}_t; \theta^l) 
\end{equation}
where $M$ denotes the set of masked indices in time domain and $\mathbf{h}_t^l$ is the $l$-th layer output for step $t$. For each layer, we use different prediction weights and different codeword embeddings. Thus, Equation (\ref{eq1}) is extended as 
\begin{equation}
    p^l(c|\mathbf{h}_t^l) = \frac{{\exp}({sim}(\mathbf{W}^l\mathbf{h}_t^l, \mathbf{e}^l_c)/\tau)}{\sum_{c'=1}^{C}{\exp}({ sim}(\mathbf{W}^l\mathbf{h}_t^l, \mathbf{e}_{c'}^l)/\tau)}
    \label{eq3}
\end{equation} 

As our method will influence the choice of layer to generate the training targets after the first iteration, we only apply intermediate layer supervision on the second iteration for a fair comparison.   Besides, the bucket relative position embedding is also applied only in the second iteration.

After pre-training, we remove all the prediction layers $\mathbf{W}^l$ and fine-tune the model with CTC loss. The CTC loss is applied only on the top Transformer layer. 
\subsection{Stabilization of Training}
Currently, it is a common practice to use 16-bit float precision (fp16) or mixed precision to pre-train large models for faster computation and less GPU memory consumption. Unfortunately, the training is unstable for large model due to the overflow issue 
(characterized by NaN losses) \cite{ding2021cogview}. A major reason is the attention score $\frac{ \mathbf{q}_i \cdot \mathbf{k}_j }{\sqrt{d}}$ is larger than the upper bound value of the fp16, resulting in the overflow issue in training. %, even if the computation order is changed to $\frac{ \mathbf{q}_i}{\sqrt{d_{k}}} \cdot \mathbf{k}_j$  

To tackle the problem, we apply a simple trick to alleviate the overflow issue. Given that softmax is invariant under translation by the same value in each coordinate that $ \text{softmax}(\mathbf{x}+ \alpha)_k = \text{softmax}(\mathbf{x} )_k$,%\footnote{\url{https://en.wikipedia.org/wiki/Softmax_function}}
%\begin{equation}
%    \text{softmax}(\mathbf{x}+ \alpha)_k = \frac{\exp(x_k+\alpha)}{\sum_k{\exp()}} = \frac{e^{x_k } e^ a }{\sum_k{e^{x_k}  e^a}} = \text{softmax}(\mathbf{x} )_k
%\end{equation}  
where $\alpha$ denotes a constant number, the equation (\ref{eq3}) can be implemented as 
\begin{equation}
\begin{split}
    \alpha_{i,j} & \propto \exp\{\frac{ \mathbf{q}_i \cdot \mathbf{k}_j }{\sqrt{d}} + r_{i-j} \} \\
    & = \exp \{(\frac{ \mathbf{q}_i}{c \sqrt{d}} \cdot \mathbf{k}_j - \max_{j' \leq T}(\frac{ \mathbf{q}_i}{c \sqrt{d}} \cdot \mathbf{k}_{j'})) \times c + r_{i-j} \}. 
\end{split}
\end{equation}
where $c$ is a scale hyperparameter and set to 32 in our work. In this way, the overflow issue could be solved, since $ \max_{j' \leq T}(\frac{ \mathbf{q}_i}{c \sqrt{d}} \cdot \mathbf{k}_{j'})$ could guarantee the max value is smaller than $2^{16}$. 
%LM FUSE VALIDATION
\iffalse
\begin{small}
\begin{table}[t]
\begin{center}
\begin{tabular}{l|ccc}
\toprule
    & 1 hour & 10 hours & 100 hours \\ \hline
update steps &  13k  & 20k  & 80k \\ 
learning rate & 5e-5 & 2e-5 & 3e-5  \\ 
freeze steps & 4k & 10k & 0  \\ 
$w_1$ & 2.90 & 2.46 & 2.15   \\
$w_2$ & -1.62 & -0.59 & -0.52 \\
\bottomrule
\end{tabular}
\caption{Fine-tuning and decoding hyperparameters.}
\label{hypers}
\end{center}
\vskip -0.3in
\end{table}
\end{small}
\fi
\label{sec:exp}

\section{Experimental Setup}
\subsection{Dataset}
For unsupervised pre-training, we use the 960-hour audio data from  LibriSpeech corpus \cite{DBLP:conf/icassp/PanayotovCPK15} to train the \textsc{Base} model and the 60,000-hour unlabeled audio from LibriLight \cite{} to train the \textsc{Large} model. For supervised fine-tuning, we consider four different partitions: Libri-light 1 hour, 10 hour splits \cite{DBLP:conf/icassp/KahnRZKXMKLCFLS20}, LibriSpeech train-clean-100 subset and LibriSpeech 960-hour full dataset. The 960-hour fine-tuning set is only used for \textsc{Large} setting. We follow the evaluation protocol of Libri-light for these splits and evaluate on the standard LibriSpeech test-clean/other sets.

\subsection{Experiment Setup}
Our pre-training process and hyperparameters follow \cite{hubert}. For the \textsc{Base} model, we first run k-means with 100 classes on 39-dimensional MFCC features over 960 hours training set to generate the discrete targets and train the first iteration model. Then we run k-means with 500 classes on the 6-th layer output features to generate the next generation targets. For both iterations, the model is trained on 32 GPUs. We crop each example to 15.6 seconds and batch the crops together to build batches not exceed 87.5 seconds. The first iteration is trained for 250k steps and the second iteration is trained for 400k steps.

For the \textsc{Large} model, we use the released HuBERT \textsc{Base} \footnote{{https://dl.fbaipublicfiles.com/hubert/hubert\_base\_ls960.pt}} and use its 9-th layer to generate target labels. The model is trained on 128 GPUs for 600k steps. The batch sizes are reduced to 53.75 seconds due to memory constraints. 

For all experiment configurations, we randomly sample the starting positions with a probability of 0.08 and mask the subsequent 10 time steps. The masked spans may overlap.   Models are optimized with AdamW optimizer with weight decay 0.01 and $\beta=(0.9,0.98)$. The learning rate ramps up linearly for the first 32k steps and then decays linearly back to 0. The peak learning rates for \textsc{Base} and \textsc{Large} are 5e-4 and 1.5e-3 respectively.
We select the intermediate layer to add supervision based on its WER on the dev set. Finally, the 4-th layer and the 9-th layer are selected for the base model and large model, respectively. 

%\subsection{Fine-tuning and Decoding}
We fine-tune the \textsc{Base} model on 8 GPUs with a batch size of 200 seconds of audio per GPU and the \textsc{Large} model on 24 GPUs with a batch size of 80 seconds of audio.  The convolutional encoder is always fixed and we use a \textit{freeze-step} hyperparameter to control how many fine-tuning steps the Transformer encoder are fixed. We optimize with Adam and a tri-stage rate schedule where the learning rate is first warmed up and held constant and then linearly decayed fofor the first 10\% of updates, held constant for the next 40\% and then linearly decayed for the remainder. For evaluation, we use wav2letter++ \cite{DBLP:journals/corr/abs-1812-07625} beam search decoder with beam size 1500 for 4-gram language model (LM) fused decoding as :
\begin{equation}
    {\rm log}p_{CTC}(\bm{y}|\bm{x}) + w_1 {\rm log} p_{LM}(\bm{y}) + w_2 |\bm{y}|
\end{equation}
The hyperparameter settings mainly follow wav2vec2.0 \cite{DBLP:conf/nips/BaevskiZMA20}. 
\begin{table}[t]
\begin{center}
\begin{small}
\begin{tabular}{lccc}
\toprule
Model & LM &   test-clean & test-other \\ \hline
\midrule
\textbf{\textit{1-hour labeled}} \\ \hline
wav2vec 2.0 \cite{DBLP:conf/nips/BaevskiZMA20} & None & 24.5 & 29.7  \\ 
HuBERT  & None   & 20.9 & 27.5  \\
WavLM \cite{chen2021wavlm}  & None &  24.5 & 29.2\\
DeCoAR 2.0 \cite{DBLP:journals/corr/abs-2012-06659} & 4-gram & 13.8 & 29.1  \\
DiscreteBERT\cite{DBLP:journals/corr/abs-1911-03912} & 4-gram & 9.0 & 17.6  \\
wav2vec 2.0 \cite{DBLP:conf/nips/BaevskiZMA20} & 4-gram  & 5.5 & 11.3 \\
%HUBERT (re-imple) & 4-gram  &  &  \\
HuBERT \cite{hubert} & 4-gram &  6.1 & 11.3\\ 
WavLM \cite{chen2021wavlm}  & 4-gram &  5.7 & 10.8\\ \hline
ILS-SSL & None &  17.9 & 23.1 \\ 
ILS-SSL &  4-gram &  5.4 &  10.2 \\ \hline
\midrule
\textbf{\textit{10-hour labeled}} \\ \hline
wav2vec 2.0 \cite{DBLP:conf/nips/BaevskiZMA20} & None  & 11.1 & 17.6 \\ 
HuBERT  & None  & 10.1 & 16.8 \\
WavLM \cite{chen2021wavlm}  & None &  9.8 & 16.0 \\
DeCoAR 2.0 \cite{DBLP:journals/corr/abs-2012-06659} & 4-gram & 5.4 & 13.3  \\
DiscreteBERT\cite{DBLP:journals/corr/abs-1911-03912} & 4-gram & 5.9 & 14.1  \\
wav2vec 2.0 \cite{DBLP:conf/nips/BaevskiZMA20} & 4-gram  & 4.3 & 9.5 \\
%HUBERT (re-imple) & 4-gram  & 4.5 & 9.5\\
HuBERT \cite{hubert} & 4-gram & 4.3 & 9.4 \\ 
WavLM \cite{chen2021wavlm}  & 4-gram &  4.3 & 9.2\\ \hline
ILS-SSL  & None & 8.3  & 13.6 \\ 
ILS-SSL &  4-gram  & 3.8 & 8.1 \\ \hline
\midrule
\textbf{\textit{100-hour labeled}} \\ \hline
wav2vec 2.0 \cite{DBLP:conf/nips/BaevskiZMA20} & None & 6.1 & 13.3 \\ 
HuBERT & None & 6.3 & 13.2 \\
WavLM \cite{chen2021wavlm}  & None &  5.7 & 12.0 \\
DeCoAR 2.0 \cite{DBLP:journals/corr/abs-2012-06659} & 4-gram & 5.0 & 12.1  \\
DiscreteBERT\cite{DBLP:journals/corr/abs-1911-03912} & 4-gram &  4.5 & 12.1 \\
wav2vec 2.0 \cite{DBLP:conf/nips/BaevskiZMA20} & 4-gram  & 3.4 & 8.0 \\
%HUBERT (re-imple) \cite{} & 4-gram &  \\
HuBERT \cite{hubert} & 4-gram  & 3.4 & 8.1\\ 
WavLM \cite{chen2021wavlm}  & 4-gram &  3.4 & 7.7\\ \hline
ILS-SSL  & None &  4.7 & 10.1 \\ 
ILS-SSL &  4-gram & 3.0 & 6.9   \\ \hline
\midrule
\end{tabular}
\caption{Model comparisons in the \textsc{Base} setting. The performance of HuBERT w/o LM is obtained by fintuning the public released model. }
\label{main result}
\end{small}
\end{center}
\vskip -0.2in
\end{table}

\subsection{Main Results}
Table \ref{main result} presents results of the \textsc{Base} setting, where the model size is less than 95M number of parameters. The pre-training dataset is LibriSpeech-960, and the fine-tuning partitions include 1-hour, 10-hour, and 100-hour subsets. 
We compare our method with several competitive self-supervised approaches in the literature, including DeCoAR 2.0 \cite{DBLP:journals/corr/abs-2012-06659}, DiscreteBERT \cite{DBLP:journals/corr/abs-1911-03912}, wav2vec 2.0 \cite{DBLP:conf/nips/BaevskiZMA20} and HuBERT \cite{hubert}.  As HuBERT only reports the WER with LM in their paper, we fine-tune their released model on different supervised sets and report the without LM results. For our method, we set the intermediate layer set $K$ as (4, 12). On test-clean set, our method reaches relative 14.4\%, 17.8\% and 25.4\% WER reductions against HuBERT baseline for different fine-tuning settings. The gain is even larger on noisy test-other set, where the relative WER reductions are 16.0\%, 19.0\% and 23.5\%. LM fusion can close the gap to some extent, but the superiority of our method also persists. On 100 hours labeled data evaluation, it outperforms HuBERT by 11.8\% and 14.8\% relatively, combining with a 4-gram language model. This indicates that our pre-training method can learn more ASR specific knowledge. 

Table \ref{large result} shows the results of \textsc{Large} setting, where the model size is around 300M number of parameters.   In this setting, we set intermediate layers $K$ as (9, 24). The superiority of our ILS-SSL method persists on Large setting, especially on noisy test-other set. On the full 960h fine-tuning setup, our simple Transformer with CTC can achieve WER 1.9/3.8, which is even slightly better than the competitive Conformer Transducer model with LSTM language model fusion, whose scores are 1.9/3.9. It also improves the HuBERT baseline by 9.5\%/11.6\% respectively, indicating the effectiveness of our method for large-scale models. Fusing with Transformer language model can further improve our model to 1.8/3.2  WER score. For other fine-tunining partitions, our method is consistently better than baselines in the w/o language model setting, and the gain becomes smaller when combining with a Transformer language model. 

%ombining with the Transformer language model, our model shows  %However, the gain for the large setting is not as large as for the base setting.  One possible explanation is xxxx... C

%From the table, we can see that our method significantly outperforms the baselines either with or without language module fusion. 

\begin{table}
\centering
\begin{small}
\begin{tabular}{lccc}
\toprule
Model & LM &   test-clean & test-other \\ \hline
\midrule
\textbf{\textit{1-hour labeled}} \\ \hline
wav2vec2.0 \cite{DBLP:conf/nips/BaevskiZMA20}  & None & 17.2 & 20.3 \\
HuBERT  & None & 17.4 & 20.3 \\
wav2vec 2.0 \cite{DBLP:conf/nips/BaevskiZMA20} & 4-gram & 3.8 & 7.1  \\ 
HuBERT & 4-gram   & 3.9 & 6.6   \\
wav2vec 2.0 \cite{DBLP:conf/nips/BaevskiZMA20} & Transf  & 2.9 & 5.8 \\
HuBERT \cite{hubert} & Transf &  2.9 & 5.4\\ \hline
ILS-SSL & None &  14.3 & 16.9 \\%17.1 & 19.7 \\ 
ILS-SSL & 4-gram &  3.6 & 6.5 \\ 
ILS-SSL &  Transf &  2.8 & 5.3 \\ \hline
\midrule
\textbf{\textit{10-hour labeled}} \\ \hline
wav2vec2.0 \cite{DBLP:conf/nips/BaevskiZMA20}  & None & 6.3 & 10.0 \\
HuBERT & None &  6.2 & 9.6 \\
wav2vec 2.0 \cite{DBLP:conf/nips/BaevskiZMA20} & 4-gram  & 3.0 & 5.8 \\ 
HuBERT & 4-gram  & 2.9 & 5.4 \\
wav2vec 2.0 \cite{DBLP:conf/nips/BaevskiZMA20} & Transf  & 2.6 & 4.9\\
HuBERT \cite{hubert} & Transf & 2.4 & 4.6\\ \hline
ILS-SSL & None & 6.1 & 9.1 \\
ILS-SSL  & 4-gram & 2.8 & 5.2 \\ 
ILS-SSL & Transf  & 2.5  & 4.5 \\ \hline
\midrule
\textbf{\textit{100-hour labeled}} \\ \hline
wav2vec2.0 \cite{DBLP:conf/nips/BaevskiZMA20}   & None & 3.1 & 6.3 \\
HuBERT  & None & 2.9 & 6.0 \\
wav2vec 2.0 \cite{DBLP:conf/nips/BaevskiZMA20} & 4-gram & 2.3  & 4.6 \\ 
HuBERT  & 4-gram & 2.3 & 4.5 \\
wav2vec 2.0 \cite{DBLP:conf/nips/BaevskiZMA20} &Transf  & 2.0 &4.0 \\
HuBERT \cite{hubert} & Transf & 2.1 & 3.9\\ \hline
ILS-SSL & None & 2.9  & 5.8 \\
ILS-SSL  & 4-gram & 2.2 & 4.5 \\ 
ILS-SSL &  Transf & 2.0 & 4.0 \\ \hline
\midrule
\textbf{\textit{960-hour labeled}} \\ \hline
Transformer-CTC \cite{synnaeve2019end} & Transf & 2.5 & 5.5 \\ 
Transformer-S2S \cite{synnaeve2019end} & Transf & 2.3 & 5.2 \\ 
Transformer-T \cite{zhang2020transformer} & Transf & 2.0 & 4.6 \\ 
Conformer-T\cite{conformer}  & LSTM & 1.9 & 3.9 \\ 
wav2vec 2.0 \cite{DBLP:conf/nips/BaevskiZMA20} & None & 2.2 & 4.5 \\ 
HuBERT  & None &  2.1 & 4.3\\
wav2vec 2.0 \cite{DBLP:conf/nips/BaevskiZMA20} & 4-gram & 2.0 & 3.6 \\ 
HuBERT  & 4-gram & 2.0 & 3.7 \\
wav2vec 2.0 \cite{DBLP:conf/nips/BaevskiZMA20} & Transf  & 1.8 & 3.3 \\
HuBERT \cite{hubert} & Transf  & 1.9 & 3.3 \\ \hline
ILS-SSL  & None &  1.9 &  3.8 \\ 
ILS-SSL & 4-gram & 1.9 & 3.4 \\
ILS-SSL &  Transf & 1.8 & 3.2  \\ \hline
\midrule
\end{tabular}
\end{small}
\caption{Model comparisons in the \textsc{Large} setting. We follow HuBERT to combine a Transformer LM in the \textsc{Large} setting. The HuBERT w/o LM and w/ 4-gram LM scores are obtained by finetuning or directly inference with the public released models\protect\footnotemark. \label{large result} }
\label{large result}
\end{table}
\footnotetext{{https://dl.fbaipublicfiles.com/hubert/hubert\_large\_ll60k.pt} and {https://dl.fbaipublicfiles.com/hubert/hubert\_large\_ll60k\_finetune\_ls960.pt}}
\subsection{Analysis}

\subsubsection{Cluster Quality Evaluation}
In this section, we analyze whether our assumption is hold that ISL-SSL can learn more SR specific information. We conduct the analysis using our pre-trained \textsc{Base} model. Following HuBERT, we obtain representations from each layer and run k-means with cluster 500 on 10\% data randomly sampled from the Librispeech training corpus.  Then we measure the correlation between the clustering assignments and the frame-level forced-aligned phonetic transcripts\footnote{{https://github.com/CorentinJ/librispeech-alignments}}. The correlation is measured by three metrics based on their co-occurrence: cluster purity, phone purity and phone normalized mutual information (PNMI) \cite{hubert}. We evaluate on the joint set of dev-clean and dev-other\footnote{The results for HuBERT evaluation is slightly different from the reported number in their paper since we use different phonetic force-alignment tool.}.  Figure \ref{analysis} shows the comparison. Different layers of our model perform similarly when aligned with phoneme units. Compared to HuBERT, ILS-SSL enforces the lower Transformer layers to learn more phonetic information while keep the capacity of the higher layers.  Block 2-12 of our model can achieve a PNMI score over 0.6, while for baseline, only block 5-12 can achieve this. This illustrates  why ILS-SSL can help for ASR task. 

%It is obviously that our model can use its more capacity to enforce the lower layer to learn more phonetic information while the capacity of high layer are ke

%Our model can achieve a high PNMI score  Even the lower layer can achieve a high Obviously, our method results in better clustering quality over all three metrics than HuBERT, especially for lower layers. Block 2-12 of our model can achieve a PNMI score over 0.6, while for baseline, only block 5-12 can achieve this. Intermediate layer supervision helps the lower layers learn more phonetic information, thus the burden on higher layers is reduced. 
\begin{small}
\begin{figure}
\centering
  \includegraphics[width=0.43\textwidth]{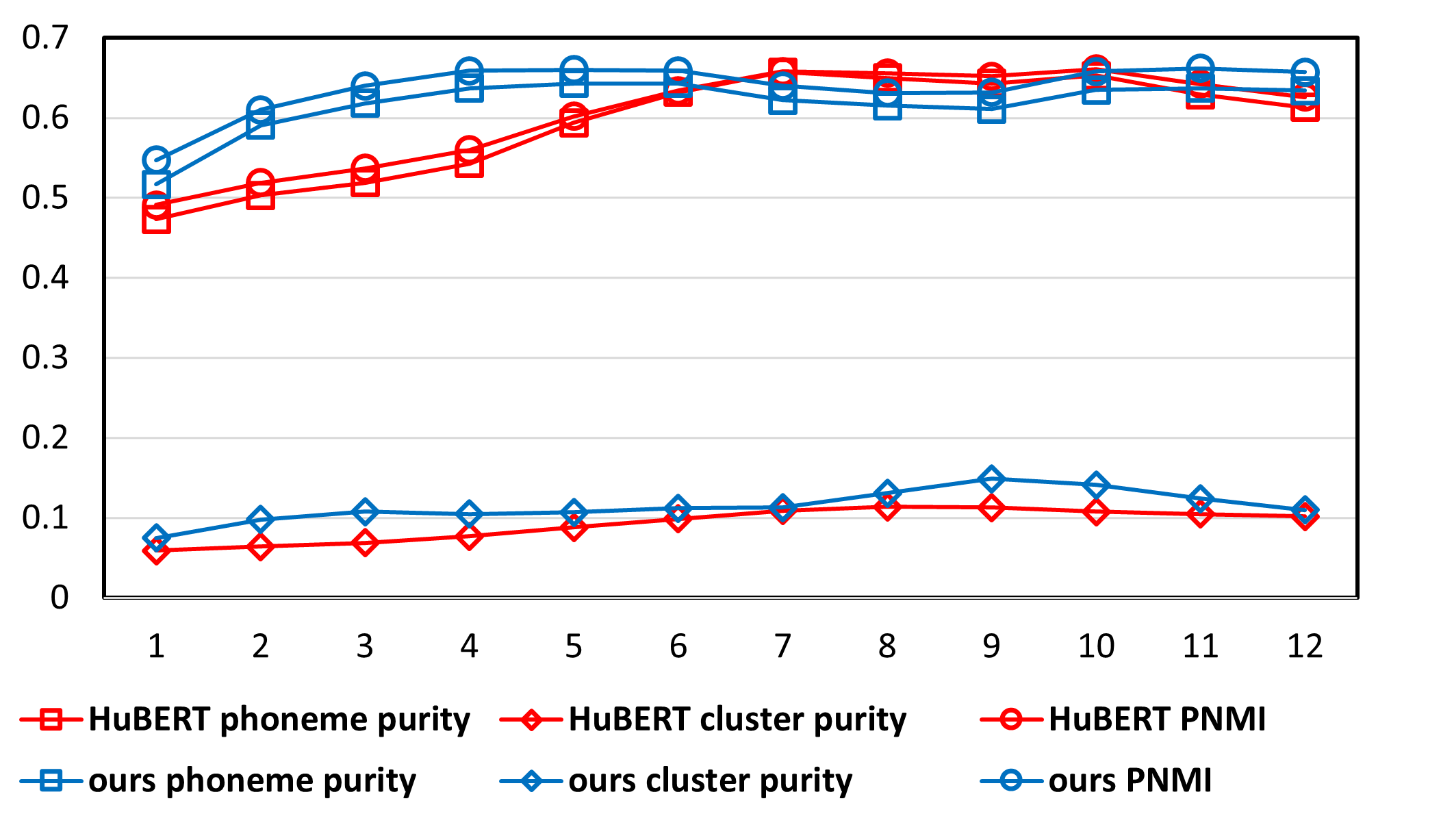}
  \caption{Quality of the cluster assignments obtained by running k-means clustering on features extracted from each Transformer layer of the \textsc{Base} model (after the 2nd iteration ). }
\label{analysis}
\end{figure}
\end{small}

\begin{table*}[h]
\centering
\resizebox{1.0\textwidth}{!}{
\begin{tabular}{l|c|c||r|r|r||r|r|r|r||r|rr||r}

\toprule
\multirow{3}{*}{Method} & \multirow{3}{*}{\#Params} & \multirow{3}{*}{Corpus}
 & \multicolumn{3}{c||}{Speaker} & \multicolumn{4}{c||}{Content} & \multicolumn{3}{c||}{Semantics}  & ParaL \\ \cline{4-14}

& & & SID & ASV & SD & PR & ASR & KS & QbE & IC & \multicolumn{2}{c||}{SF} & ER  \\ \cline{4-14}

& & & Acc $\uparrow$ & EER $\downarrow$ & DER $\downarrow$ & PER $\downarrow$ & WER $\downarrow$ & Acc $\uparrow$ & MTWV $\uparrow$  & Acc $\uparrow$ & F1 $\uparrow$ & CER $\downarrow$ & Acc $\uparrow$ \\ \hline 

FBANK & 0 & - & 8.5E-4 & 9.56 & 10.05 & 82.01 & 23.18  & 8.63 & 0.0058 & 9.10 & 69.64 & 52.94 & 35.39 \\ \hline

HuBERT \textsc{Base} ~\cite{hubert} & 94.68M & LS 960 hr &  \textbf{81.42} &  \textbf{5.11} &  \textbf{5.88}  & 5.41 & 6.42 & 96.30 & 0.0736 & 98.34 & 88.53 & 25.20 & 64.92 \\
ILS-SSL \textsc{Base} & 94.68M & LS 960 hr & 79.29 & 5.24 & 6.31 &  \textbf{5} & \textbf{5.45} & \textbf{4.38} &  \textbf{0.0789} &  \textbf{98.47} &  \textbf{89.16} &  \textbf{24.29} &  \textbf{65.79} \\ \bottomrule

%WavLM Base ~\cite{}  &   94.70M & LS 960 hr & \textbf{84.51}&	\textbf{4.69}&	\textbf{4.83}&	\textbf{4.84}&	6.21	&	96.79&	\textbf{0.0870}&	\textbf{98.63}&	\textbf{89.38}&	\textbf{22.86} & \textbf{65.94}&81.9 \\ \hline
\end{tabular}}

% \vspace*{1mm}
\caption{
Universal speech representation evaluation on SUPERB benchmark. ParaL denote Paralinguistics aspect of speech.  \label{table:superb}
}
\end{table*}
\subsubsection{Evaluation on non-ASR Tasks}
Recent work shows \cite{chen2021wavlm} that speech pre-trained  models can solve full stack speech processing tasks, because the model utilizes bottom layers to learn speaker related information and top layers to encode content related information.  We evaluate the proposed method on the SUPERB benchmark \cite{superb}, which is designed to provide 
a standard and comprehensive testbed for pre-trained models on various speech tasks. It covers ten tasks, including Speaker Identification (SID),	 Automatic Speaker Verification (ASV), Speaker Diarization (SD), Phoneme Recognition (PR), Automatic Speech Recognition (ASR), Keyword Spotting (KS), Query by Example Spoken Term Detection (QbE), Intent Classification (IC), Slot Filling (SF), Emotion Recognition (ER). These tasks can be grouped into four aspects
of speech: content, speaker, semantics, and paralinguistics.

Table \ref{table:superb} compares ILS-SSL \textsc{Base} and HuBERT \textsc{Base} on SUPERB, indicating the ILS-SSL is better than HuBERT on content and semantic related tasks, while the performance degradation is observed for the speaker related tasks. An explanation is that the pseudo-labels are highly correlated with the phone information, which may bias the model to discard the speaker characteristic and focus on the phoneme recognition.

\subsubsection{Ablation Study}
In this section, we perform some ablation studies. The experiments are conducted on 10 hours labeled data using \textsc{Base} model and evaluated on the dev-other set. We show the results both with or without LM fusion. For LM fusion evaluation, we set language model weight $w_1$ as 2, insersion score $w_2$ as -1.0 and beam size as 50. The baseline HuBERT results are obtained using their released model.

We first study the effect of choices for supervised layers $K$. The results are shown in Table \ref{layer choice}. Firstly, we set the size of $K$ as 2 and the 12-th layer is always included. As shown in the table, ILS-SSL always outperforms HuBERT no matter which intermediate layer is chosen. The performance is best for setting $K=(4, 12)$. This is consistent with the our previous observation that the lower 4 layers of HuBERT obtain PNMI score lower than 0.6. Then we set the size of $K$ as 3 and evaluated on two settings: $(4, 8, 12)$ and $(6, 9, 12)$. There is no improvement compared with the setting $(4, 12)$, indicating that the more intermediate layers with supervision signal doesn't mean the better ASR performance. 

\begin{table}[h]
\begin{center}
\begin{small}
\begin{tabular}{l|cc}
\toprule
Model & w/o LM & with LM \\ \hline
HuBERT &    16.8 &  9.4      \\  \hline
ILS-SSL (2, 12)  &  14.4  & 9.0  \\ 
ILS-SSL (4, 12)  & \textbf{13.1}  & \textbf{8.3}  \\ 
ILS-SSL (6, 12)  & 13.5  &  8.7 \\
ILS-SSL (8, 12)  & 13.8  & 8.9  \\
ILS-SSL (10, 12)  & 14.7  &  9.4  \\ \hline
ILS-SSL (4, 8, 12)  & 13.4  & 8.7  \\
ILS-SSL (6, 9, 12)  & 13.3 & 8.2 \\  \hline
\bottomrule
\end{tabular}
\caption{Results on dev-other set with supervision on different intermediate layers.}
\label{layer choice}
\end{small}
\end{center}
\end{table}

\begin{table}[h]
\begin{center}
\begin{small}
\begin{tabular}{l|cc}
\toprule
Model & w/o LM & with LM \\ \hline
HuBERT &   16.8  &  9.4     \\ 
\,\,\,\,+ILS-FT (share) & 17.9  & 10.5 \\ 
\,\,\,\,+ILS-FT (sep) & 18.3 &  10.8  \\ \hline
ILS-HuBERT &  13.1 & 8.3  \\
\,\,\,\,+share IL weights & 13.5   & 8.5 \\
\,\,\,\,+ILS-FT (share) & 16.7 & 10.8 \\ 
\,\,\,\,+ILS-FT (sep) & 17.1 & 10.8 \\ 
%\,\,\,\,+larger predict layer &  &  \\ 
\,\,\,\,-bucket rel pos & 14.5  & 8.8  \\
\bottomrule
\end{tabular}
\caption{Ablation study results on dev-other set.}
\label{ablation}
\end{small}
\end{center}
\vskip -0.2in
\end{table}
Then we perform more ablation experiments with $K=(4, 12)$ and show the results in Table \ref{ablation}.  First, we share the projection weights $W^l$ and the codeword embedding $e^l_c$ in equation (\ref{eq3}) for each supervised layer. The experiment is denoted as ``+ share IL weights". The results are slightly worse but the gap is less than  3\%. This indicates the gain is come from intermediate layer supervision instead of larger pre-training model size. Next, we apply the intermediate layer supervision in fine-tuning stage (ILS-FT), where we compute CTC loss on every $l \in K$. We use either shared or separated CTC weights. The results show that the ILS method cannot generalized from SSL to supervised task. We further remove the bucket relative position embedding in our model and the performance drops a little bit, indicating the bucket relative position embedding is a supplement to convolutional position embedding.

\section{Conclusion}
In this paper, we introduce ILS-SSL method to learn a speech recognition oriented self-supervised model. The method is simple but effective. It outperforms competitive baselines on LibriSpeech benchmark in different low-resource setting. Analysis shows our method can learn representations with a higher correlation with phonetic units. In the future, we will consider how to jointly pre-train text and audio in a unified model. 


\bibliographystyle{IEEEbib}
\bibliography{strings,refs}
\end{document}